\documentclass[aip,graphicx,amsmath]{revtex4-1}
\usepackage{amssymb}
\usepackage{graphicx}
\usepackage{amsmath}
\draft 

\begin{document}


\title{On Genus-Two Solutions for the ILW equation} 



\author{Y. Tutiya}
\affiliation{ 
Kanagawa Institute of Technology 
}


\date{\today}

\begin{abstract}
The existence of theta function solutions of genus two for the ILW equation is established.\ A numerical example is also presented.\  
The method basically goes along with the Krichever's construction of theta function solutions for soliton equations, such as the KP equation.\ 
This idea leads us to a question whether a Riemann surface exists which allows a peculiar Abelian integral of the third kind.\ The answer is affirmative at least for genus-two curves.
\end{abstract}


\maketitle 

\section{Abstract}
The intermediate long-wave (ILW) equation describes the propagation of long internal gravity waves in a stratified fluid of finite depth~\cite{J77,JE78,KK78}. One of the standard dimensionless forms is as follows~\cite{NM80,P92}.\
\begin{eqnarray}
\begin{array}{l}
\displaystyle u_t+2uu_x+G[u]_{xx}=0,\\
\displaystyle G[u](x)=\lambda\int_{-\infty}^{\infty}\hspace{-1.9em}-\hspace{2em}\dfrac{u(x',t){\rm sgn}(x'-x)}{{\rm exp}(\pi \lambda|x'-x|)-1}dx'\ .
\end{array}\label{ILW-b}
\end{eqnarray}
Here, the slashed integral denotes the Cauchy's principal value of integral.
The constant \(\lambda\) characterizes the relative depths of the two fluid layers.\ 
This study aims to find theta function solutions for the ILW equation (\ref{ILW-b}).\ 
In the context of soliton theory, the phrase ``theta function solutions" indicates peculiar families of solutions parametrized by compact Riemann surfaces.\
Such solutions have already been discovered for most of celebrated soliton equations.  
However, as for nonlocal soliton equations such as the ILW equation or the intermediate nonlinear Schr\"{o}dinger equation, the only theta function solutions previously known  were of a genus one type, i.e. elliptic solutions.\
Moreover, the construction of these elliptic solutions are heavily reliant upon properties specific to elliptic curves, making it difficult to find any hint of the existence of a greater-genus solutions.\ 
This paper explains the existence of theta function solutions of genus two for the ILW equation.\
The method takes the well-known approach of constructing a Baker-Akhiezer function which satisfies the Lax system --- known as Krichever's construction.\ 
Two ideas will be important in applying this satndard technique.\ 
One is to write \(u\) as a difference of two holomorphic functions.\ An explanation of how this is achieved can be found in section 2 of this paper.\ 
The other is to see the equation (\ref{ILW-b}) as a transcendental reduction of a higher dimensional integrable system.\ This higher dimensional system is introduced in section 3 and its theta function solutions exhibited in section 4.\  
The most difficult task is to find the theta function solutions which survive reduction to the ILW setting.\ 
This raises the question as to whether a Riemann surface exists which possesses a certain Abelian integral of the third kind.\
It is proved in section 5 that such genus-two Riemann surfaces exist.\ 
Since it is represented by a kind of transcendental equation, 
one can only show the existence of a solution and can not describe it explicitly using known special functions.\ 
However, the method is nevertheless constructive, and the form of the solution is almost explicit. In section 6, a numerical example of the solution is given.\ 
Section 7 is devoted to further discussion.\
\section{Suitable differential-difference form}
\begin{figure}
\includegraphics[width=.6\linewidth]{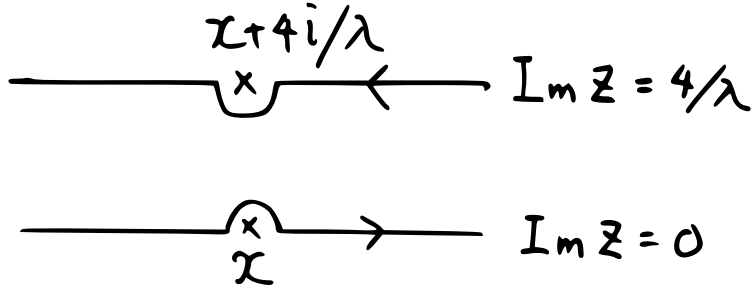}
\caption{\label{fig:contS} The integration contour \(S\).}
\end{figure}

The derivation of the ILW equation from the basic fluid dynamical equations assumes that \(u\) vanishes at \(|x|\to\infty\).\ Under this boundary condition, multi-soliton solutions were discovered by Joseph~\cite{J77,JE78}  and , subsequently obtained in other several ways by Chen and Lee~\cite{CL79} and Matsuno~\cite{M79} and Kodama {\it et al }\ \cite{KAS82}.\ 
The spatially periodic boundary condition has also been studied intensively.\ 
Although with the original derivation of the equation we might not expect a periodic solution, 
the numerical work of Kubota {\it et al} first suggested that such solutions exist~\cite{KK78}.
After this, precise expressions for periodic solutions were found~\cite{Zai83,Mil90,P92}.
In these preceding studies, the ILW equation is always transformed into a differential-difference equation so as to avoid handling the singular integral term directly~\cite{CL79,NM80,AFJS82}.\
Namely, the transformed equation contains both differentials and differences about \(x\) and does not contain any singular integral terms.\ 
If we try to rewrite (\ref{ILW-b}) as such a differential-difference equation, there is no chance to reutilize the technique for the spatially decaying boundary condition or periodic boundary condition, since theta function solutions of genus two are neither decaying or periodic.\
However, it is real analytic and bounded on the entire real axis.\
Therefore, the following lemma provides the key to introducing a complex difference.\\
{\bf Lemma 1}\\
Let \(f(x)\) be a real analytic function and be bounded on the real axis. Then, the following holds.\
\begin{eqnarray}
G[f_x](x)-\lambda f(x)=\dfrac{\lambda}{2}\int_{-\infty}^{\infty}\hspace{-1.9em}-\hspace{1.4em}\left(
\dfrac{f_x(x')}{\sinh(\pi\lambda(x'-x)/2)}-\dfrac{\pi\lambda f(x')}{4\cosh^2(\pi\lambda(x'-x)/4)}
\right)dx'\label{lem001}
\end{eqnarray}
\\[.2em]
{\it Proof.} 
Since \(f\) is bounded, it is expressible as follows by means of the integration by parts formula.\
\begin{eqnarray}
f(x)&=&\lim_{\varepsilon\to0}\left\{\left.\dfrac{f(x')}{1+\exp(-\pi\lambda(x'-x)/2)}\right|_{x'=-\infty}^{x'=x-\varepsilon}+\left.\dfrac{-f(x')}{1+\exp(\pi\lambda(x'-x)/2)}\right|_{x'=x+\varepsilon}^{x'=\infty}\right\}\nonumber\\
&=&
\int_{-\infty}^{\infty}
\dfrac{\pi\lambda f(x')dx'}{8\cosh^2(\pi\lambda(x'-x)/4)}\nonumber\\
&&\hspace{2em}+\lim_{\varepsilon\to0}\left\{\int_{-\infty}^{x-\varepsilon}
\dfrac{f_x(x')dx'}{1+\exp(-\pi\lambda(x'-x)/2)}
+\int_{x+\varepsilon}^{\infty}\dfrac{-f_x(x')dx'}{1+\exp(\pi\lambda(x'-x)/2)}
\right\}\nonumber
\end{eqnarray}
Substituting the above into the second term of the l.h.s. of (\ref{lem001}), one arrives at the r.h.s.
\hfill 
\fbox{\rule{0em}{.6em}}\\[.2em]
Next, let us introduce the following complex valued function.\
\begin{eqnarray}
A(z,t):=\dfrac{\lambda}{2}\int_{-\infty}^{\infty}\left(\dfrac{u(x',t)}{\sinh(\pi\lambda(x'-z)/2)}-\dfrac{\pi\lambda (\int_\alpha^{x'}u(\xi,t)d\xi)}{4\cosh^2(\pi\lambda(x'-z)/4)}\right)dx'\ .
\end{eqnarray}
Here, \(\alpha\) 
is an arbitrarily fixed constant.\ 
Then, we denote the analytic continuation of this function to the real axis from the upper (lower) half plain as,
\begin{eqnarray}
\displaystyle A^{\pm}(x,t):=\lim_{{\rm Im}(z)\to \pm 0}A(z,t)
=\pm iu(x,t)+G[u](x,t)-\lambda \int_{\alpha}^{x}u(\xi,t)d\xi,\quad ({\rm Re}(z)=x).
\end{eqnarray}
This shows that the following properties hold.\\
{\bf Remark 2}\\
1.\
\(A^+(x+4i/\lambda)=A^-(x)\)\\
2.\
\(A^+\) is holomorphic in the strip \(0<{\rm Im}(z)<4/\lambda\) and 
\(A^-\) is holomorphic in the strip \(-4/\lambda<{\rm Im}(z)<0\).\\[.2\baselineskip]
Conversely, we can construct \(u\) from \(A^{\pm}\) by virtue of the following lemma.\\[.2em]
{\bf Lemma 3}\\
Let \(f^+(z)\) be a complex function  holomorphic in the strip \(0<{\rm Im}(z)<4/\lambda\) and  \(f^-(z):=f^+(z+4i/\lambda)\).\ 
Suppose both \(f^{\pm}(z)\) and \({f}_z^{\pm}(z)\) are bounded when \({\rm Re}(z)\to \pm \infty\).\
Then, the following holds.\
\begin{eqnarray}
\displaystyle f_z^+(x)+f_z^-(x)=iG[f_z^-(x)-f_z^+(x)]-i\lambda(f^-(x)-f^+(x)),\quad x\in \mathbb{R}\ .\label{converse}
\end{eqnarray}
{\it Proof.}\ 
Let \(S\) be an integration contour shown in FIG.~\ref{fig:contS}.\ 
Then, with the help of lemma 1, we can consider the following calculation, 
\begin{eqnarray}
0&=&\dfrac{\lambda}{2}\int_S\left(\dfrac{f^+_x(x',t)}{\sinh(\pi\lambda(x'-z)/2)}-\dfrac{\pi\lambda f^+(x',t)}{4\cosh^2(\pi\lambda(x'-z)/4)}\right)dx'\nonumber\\[.3em]
&=&G[f^+_x]-\lambda f^+-if^+_x
-G[f^-_x]+\lambda f^--if^-_x\, ,\nonumber
\end{eqnarray}
which establishes the formula (\ref{converse}).\hfill 
\fbox{\rule{0em}{.6em}}\\[.2em]

Making use of this lemma, the equation (\ref{ILW-b}) can be rewritten into the following differential equation with a complex difference in the variable \(x\).\
\begin{eqnarray}
i(A^--A^+)_t-i\lambda(A^--A^+)_x-(A^--A^+)(A^--A^+)_x+(A^-+A^+)_{xx}=0\label{ILW-d}
\end{eqnarray}
It should be emphasized again that, if one finds a solution \(A^{\pm}\) for (\ref{ILW-d}) which satisfies the assumptions of the lemma 3, then \(u(x)=\dfrac{i}{2}(A^--A^+)\) becomes a solution for the equation (\ref{ILW-b}).\
This was the strategy  used to solve the equation (\ref{ILW-b}).\
\section{Auxiliary Lax system}
The equation (\ref{ILW-d}) is usually called \(1+1\)-dimensional because it contains one spatial variable \(x\) and one time variable \(t\).\ 
But, it is sometimes revealing to regard it as a reduction from the following \(2+1\)-dimensional equation that contains another spatial independent variable, specifically, \(r\).\ 
\begin{eqnarray}
i(\tilde{A}-A)_t
-(\tilde{A}-A+c)(\tilde{A}-A)_x+(\tilde{A}+A)_{xx}=0\ .\label{auxnlin}
\end{eqnarray}
Here, \(c\) is a certain complex constant.\ 
It looks very much like the equation (\ref{ILW-d}) but this time \(\tilde{A}\) denotes a shift in the \(r\)-direction, \(\tilde{A}(x,t,r):=A(x,t,r+\Delta)\), where \(\Delta\) is a complex constant.\ 
In this section, a Lax system associated with this \(2+1\)-dimensional equation is presented.
The strategy is that, if one finds a solution for (\ref{auxnlin}) whose \(r\)-dependence is of the form \(A(t,x+4i r/\lambda\Delta)\), it also becomes a solution for (\ref{ILW-d}) as well, because the \(r\)-shift can take the place of the \(x\)-shift.\ 
It should be noted that the notion of regarding the ILW equation as a reduction of a larger differential-difference system is already seen in the work of T and Satsuma~\cite{TS03}.\ 
Now let us introduce the Lax equation for (\ref{auxnlin}) as follows.
\begin{eqnarray}
\left\{\begin{array}{l}
\kappa\tilde{\Psi }=(\partial + U)\Psi \\
i\Psi _t=(\partial^2+V)\Psi 
\end{array}\right.\ , \label{auxLax}
\end{eqnarray}
where \(\kappa\) is a purely imaginary constant and \(\partial\) denotes \(\partial/\partial x\). 
Then, the compatibility condition between the evolution in \(t\)-direction and that of \(r\)-direction becomes
\begin{eqnarray}
\left\{\begin{array}{l}
\tilde{V}-V=-2U_x\\
iU_t+2UU_x+\dfrac{1}{2}(\tilde{V}+V)_{x}=0
\end{array}\right.\ .\label{auxnonlin}
\end{eqnarray}
By setting \(A=-\int Vdx/2\), the first equation in (\ref{auxnonlin}) becomes \(U=\tilde{A}-A+c\) where \(c\) is a constant and the second equation becomes nothing but (\ref{auxnlin}).\ 
It should be remarked that, if one replaces the \(r\)-shifts by the \(x\)-shifts in the equation (\ref{auxLax}), it becomes the Lax equation for the ILW equation reported in the work of Kodama {\it et al}.~\cite{KSA81}\ 
\section{Theta function solutions for the auxiliary Lax system}
In this section, theta function solutions of arbitrary genus for the equation (\ref{auxnlin}) will be presented. 
The final purpose of this paper is solely to establish a solution of genus two for the ILW equation.\
But it will give a better perspective to see the solutions for (\ref{auxnonlin})  in this wider setting.\
Enormous effort went towards finding theta function solutions for soliton equations around 1975, thanks to which, it is a relatively straightforward exercise these days to find solutions to equations like (\ref{auxLax}).\ We shall skip further historical remarks and just list [\onlinecite{Ak61,Dub75,Dub75-2,IM75,IM75-2,MM75,Kri76,Kri77,Kri81}] as references instead.\ 
In this paper, the explanations and notations about this topic basically follows those in the section 2 and 3 of the textbook [\onlinecite{BBEIM94}].\\[.2em] 
{\bf Definition 4}\\
Let \(X\) be a compact Riemann surface.\
We will use the following notations about \(X\):\\[.3\baselineskip]
\(a_1,\cdots,a_g,b_1,\cdots,b_g\) are canonical homology cycles on \(X\).\\[.3\baselineskip]
\(\omega_1,\cdots,\omega_g\) are a set of normalized differentials of the first kind that satisfy
 \(
\int_{a_j}\omega_k=2\pi i\delta_{jk}
\).\\[.3\baselineskip]
\(P_{0}\) and \(P_*\) are distinct two points on \(X\).\\[.3\baselineskip]
\(\chi_{*}\) is an Abelian differential of the third kind which  is holomorphic in \(X\backslash\{P_0, P_*\}\) and possesses a pole of degree 1 at \(P_*\) with the residue \(+1\) and at \( P_0\) with the residue \(-1\).\ 
Moreover, all the \(a_j\)-periods of \(\chi_{*}\) vanish.\\[.3\baselineskip]
\(z\) is a local coordinate around \(P_0\) which satisfies \(z(P)=0\).\\[.3\baselineskip]
\(\chi_{j},\ (j=1,2)\) is a normalized differential of the second kind with a single pole at \(P_0\) and whose expansion in the neighborhood of \(P_0\) is 
\((-z^{-j-1}+O(1))dz\).\ \(\chi_j\) is holomorphic in \(X\) except for \(P_0\) and all its \(a\)-periods are zero.\\[.3\baselineskip]
\(B\) is the period matrix and \(
(B)_{ij}=\int_{b_i}\omega_j
\).\\
\(\Theta\) is the Riemann's theta function of genus \(g\) defined as;
\begin{eqnarray}
\Theta(\vec{z};B)
=\sum_{m\in \mathbb{Z}^g}\exp\left\{
\frac12\vec{m}^t\ B\ \vec{m}+\vec{m}^t\ \vec{z}
\right\}.
\end{eqnarray}
\(\vec{A}\) is the Abel-Jacobi map;
\begin{eqnarray}
&&\vec{A}(P)=
\left(\cdots,\int_{P_0}^P\omega_j,\ \cdots\right)^t.
\end{eqnarray}
The \(b_j\)-cycles of these differentials will be denoted in vector forms as;
\begin{eqnarray}
&&\vec{U}_*=
\left(\cdots,\int_{b_j}\chi_{ *},\ 
\cdots\right)^t,\quad
\vec{U}_\alpha=
\left(\cdots,\int_{b_j}\chi_{\alpha},\ \cdots\right)^t,\quad (\alpha=1,2)\ .
\end{eqnarray}
\(Q_*\)  and \(Q_j, \ (j=1,2)\) are defined to satisfy;
\begin{eqnarray}
&&\int_{Q_*}^P\chi_{*}
=-\log z+o(z^1)\ ,\quad 
\int_{Q_j}^P\chi_{j}
=z^{-j}+o(z^1)\ .
\label{Qcond}
\end{eqnarray}
\quad\\
This setup allows us to introduce a Baker-Akhiezer function in the following form.
\begin{eqnarray}
&&\Psi(x,t,r,P):=\frac{
\Theta(\vec{A}(P)+\kappa x\vec{U}_1
-i\kappa^2t\vec{U}_2+\Delta^{-1}r\vec{U}_*+\vec{Z})
\Theta(\vec{Z})
}{
\Theta(\vec{A}(P)+\vec{Z})
\Theta(\kappa x\vec{U}_1
-i\kappa^2t\vec{U}_2+\Delta^{-1}r\vec{U}_*+\vec{Z})
}\cdot\nonumber\\
&&\hspace*{10em}\cdot
\exp\big(
\kappa  x\int_{Q_1}^P \chi_1
-i\kappa^2t\int_{Q_2}^P \chi_2
+\frac{r}{\Delta}\int_{Q_*}^P\chi_*
\big)\ ,\label{Psi}
\end{eqnarray}
where \(\vec{Z}\) is a real constant vector chosen to ensure that \(\Theta(\vec{A}(P)+\vec{Z})\) is not identically zero.\\
\(\Psi(P)\) is logarithmically ramified at the points \(P_*\) and \(P_0\).\ We so set the branch cut between \(P_*\) and \(P_0\) as not to cross any of \(a_j\) or \(b_j\) \((j=1,2,\cdots, g)\).\ 
Then, with a straightforward calculation, it is noticable that \(\Psi(P)\) is invariant when the point \(P\) goes through one full revolution on any cycle of \(a_j\) or \(b_j\) \((j=1,2,\cdots,g)\).\
In the vicinity of \(P_0\), \(\Psi(P)\) is representable as 
\begin{eqnarray}
&&\Psi(P)=
(1+f(r,x,t)z+\cdots)
z^{-r/\Delta}\exp\big(
\kappa xz^{-1}-i\kappa^2tz^{-2}\big)\ .
\end{eqnarray}
Hence, \((i\Psi_t-\Psi_{xx})/\Psi\) and \((\kappa\tilde{\Psi}-\Psi_x)/\Psi\) are holomorphic around \(P_0\).\ 
We also see that \(\Psi_t/\Psi\), \(\Psi_x/\Psi\), \(\Psi_{xx}\) and \(\tilde{\Psi}/\Psi\) are also holomorphic aroud \(P_*\), since the \(r\)-shift only increase the degree of the zero of \(\Psi\) at \(P_*\).\  
Hence, \((i\Psi_t-\Psi_{xx})/\Psi\) and \((\kappa\tilde{\Psi}-\Psi_x)/\Psi\) are also holomorphic around \(P_*\).\ 
Hence, these two functions are meromorphic functions whose \(g\) zeros are the same as those of  
\begin{eqnarray}
\Theta(\vec{A}(P)+\kappa  x\vec{U}_1
+i\kappa^2t\vec{U}_2+\Delta^{-1}r\vec{U}_*+\vec{Z})\ .
\nonumber\end{eqnarray}
But such a meromorphic function can not exist except for constant functions since those \(g\) zeros are located at general positions.\
Hence, 
\begin{eqnarray}
U=(\kappa\Psi(x,t,r+\Delta)-\Psi_x)/\Psi
,\quad V=(\Psi_t-\Psi_{xx})/\Psi\label{UV}
\end{eqnarray}
do not dependo on \(P\).\ 
These are exactly the same as the auxiliary Lax system (\ref{auxLax}).\\[.2em] 
{\bf Proposition 5}\\
\(\Psi\) in (\ref{Psi}) is a solution for the Lax system (\ref{auxLax}) and, \(U\) and \(V\) in (\ref{UV}) are solutions for the compatibility condition (\ref{auxnonlin}).
\section{Theta function solution for ILW of genus 2}
\begin{figure}
\includegraphics[width=.6\linewidth]{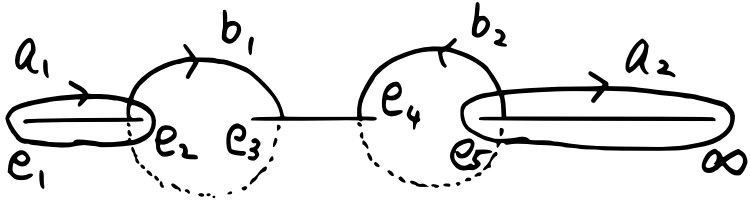}
\caption{\label{Fig2} The homology cycles for theorem 6.}
\end{figure}

\begin{figure}
\includegraphics[width=.5\linewidth]{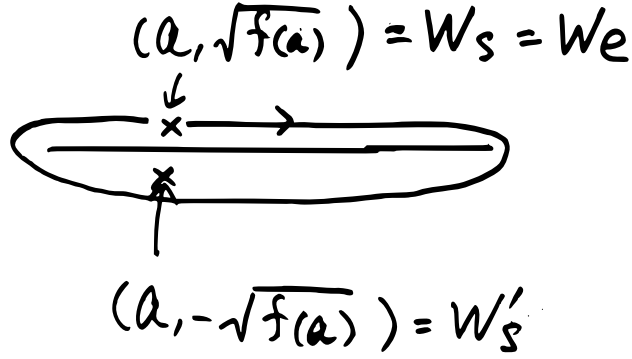}
\caption{\label{Fig3} An illustration of the segment \({\cal W}\).}
\end{figure}

\noindent {\bf Theorem 6}\\
There exist combinations of a Riemann surface \(X\) of genus two, a set of homology cycles \(\{a_1,a_2,b_1,b_2\}\) and distinct two points \(P_0, P_*\) which have the following properties:
\begin{itemize}
\item[1.]
The components of the matrix \(B\) are real.\ 
\item[2.]
\(\vec{U}_1, \vec{U}_2\) and \(\vec{U}_*\) are purely imaginary vectors.\  
\item[3.]
There exixts a real nonzero constant \(\gamma\) which satisfies
\begin{eqnarray}
\vec{U}_*=\gamma\times \vec{U}_1\ .
\label{redcond}
\end{eqnarray}
\end{itemize}
This is the main theorem of this paper since it will immediately give a theta function solution for the ILW equation (\ref{ILW-b}) with the corollary below.\\[.2em] 
{\bf Corollary 7}\quad (The existence of genus-two solutions for ILW equation)\\
If we chose \(X, P_0, P_*\) and homology cycles that make the theorem hold, then, \(u=iU-2i\kappa/\gamma\) satisfies the ILW equation (\ref{ILW-b}) with \(\lambda=4i\kappa/\gamma.\)\\
{\it Proof of Corollary 7}.\\
Since \(U\) does not depend on \(P\) and \(\tilde{\Psi}/\Psi|_{P=P_*}=0\), we set \(P=P_*\) so that \(U\) becomes expressible as;
\begin{eqnarray}
\begin{array}{l}
\displaystyle U=-\dfrac{F^-_x}{F^-}+\dfrac{F^+_x}{F^-}-\kappa\int_{Q_1}^{P_*}\chi_{1},\\
\displaystyle F(x,t):=\Theta(\kappa  x\vec{U}_1
-i\kappa^2t\vec{U}_2+\Delta^{-1}r\vec{U}_*+\vec{Z}),\quad F^\pm:=F(x\mp\gamma/2\kappa,t).
\end{array}\label{sol-U}
\end{eqnarray} 
Now, one can readily identify \(-F^+_x/F^+\) as \(A\), \(-F^-_x/F^-\) as \(\tilde{A}\), and \(-\kappa\int_{Q_1}^{P_*}\chi_1\) as \(c\) in 
the formula (\ref{auxnlin}).\
Suppose \(\kappa\) and \(\Delta\) are purely imaginary and \(\vec{Z}\) is real.\ 
Then, \(F(x,t)\) is real-valued and has no zero for \(x,t\in \mathbb{R}\) due to the same argument as in [BBEIM, p.63].\ Namely, \(F(x,t)=1+\sum_{m_1\geq0,m_2\in\mathbb{Z}}\exp(\vec{m}^tB\vec{m})
\cosh(\vec{m}^t(\kappa  x\vec{U}_1
-i\kappa^2t\vec{U}_2+\Delta^{-1}r\vec{U}_*+\vec{Z}))>0\). 
 Hence, \(U\) is holomorphic in the strip \(-\gamma/2\kappa<{\rm Im}(x)<\gamma/2\kappa\) if \(|{\rm Im}(\gamma/2\kappa)|\) is sufficiently small (or, in other words, \(|\vec{U}_*|\) is sufficiently small).\ 
By virtue of  Lemma 3, \(U\) now satisfies 
\begin{eqnarray}
iU_t-(U-4\kappa/\gamma)U_x+iG[U]_{xx}=0\ .
\end{eqnarray}
Hence, 
\begin{eqnarray}
u=i(U/2-2\kappa/\gamma)\label{sol-b}
\end{eqnarray}
satisfies the original ILW equation (\ref{ILW-b}) for the relative depth \(\lambda=4i\kappa/\gamma\).
 \hfill({\it The proof of the corollary 7 is finished.})\\[.2\baselineskip]
Now let us  proceed to prove theorem 6.\\[.2\baselineskip]
{\it Proof of  Theorem 6}.\quad
We will explicitly construct an example of such a Riemann surface.\ Let \(e_1<e_2<e_3<e_4<e_5\) be real numbers.\
Let \(C\) be the hyperelliptic curve compactifying an affine curve \(t^2=f(s)=\prod_{j=1}^5(s-e_j)\).\ 
Let us set the representatives of the canonical homology cycles as 
shown in FIG.\ref{Fig2}.\\
\noindent The following lemma is crucial.\\[.2\baselineskip]
{\bf Lemma 8}\\
There exist a pair of real numbers \(a\) and \(c\) such that \(a,c>e_5\) and \(\displaystyle \int_a^c\dfrac{s-a}{\sqrt{f(s)}}ds=0\).\ 
Here, the integration contour starts at \((a,{f(a)}^{1/2})\), passes through \(\infty\) and ends at \((c,{f(c)}^{1/2})\).\\[.5\baselineskip]
{\it Proof of lemma 8.}\quad  
Let \({\cal W}\) be a segment which starts at \((a,\sqrt{f(a)})\), passes through \(\infty\), \((e_5,0)\), and comes back to \((a,\sqrt{f(a)})\) (FIG.\ref{Fig3}).\\
 We temporarily distinguish the starting point and the end point,  and denote them as \(W_s\) and \(W_e\) respectively.\
Next, we consider \(\Omega:{\cal W}\longrightarrow \mathbb{R}\);
\[
(w,q)\mapsto \int_{a}^{w}\dfrac{(s-a)ds}{\sqrt{f(s)}}\ ,
\]
where the integration contour starts at \(W_s\) and lies on \({\cal W}\).
Suppose \(\Omega(W_e)\) is negative.\ 
We notice that \(\Omega\) is analytic and \(\Omega(W_s)=0\).\
Moreover it becomes locally minimum only at \(W_s\) and locally maximum at \({W'}_s:=(a,-\sqrt{f(s)})\).\ 
Hence, there must be a point \((c,f(c)^{1/2})\in {\cal W}\) such that \(\Omega((c,f(c)^{1/2}))=0\).\
If \(\Omega(W_e)\) is positive, we redefine \({\cal W}\) to be a segment which starts at \(W_s:=(a,-\sqrt{f(a)})\), passes through \((e_5,0)\), \(\infty\), and comes back to \(W_e=(a,-\sqrt{f(a)})\).\ 
Then, the stationary points again tell us that \(\Omega\) vanishes at a certain point \((c,f(c)^{1/2})\).\ 
If \(\Omega(W_e)\) is zero, we move \(a\) slightly so that it becomes nonzero.\\ 
\hfill ({\it The proof of lemma 8 is finished.})\\
Let us come back to the proof of theorem 6.\
We set real numbers \(a\) and \(c\) so as to make lemma 8 holds.\
Then, we introduce a set of Abelian differentials of the first kind as 
\begin{eqnarray}
\hat{\omega}_1=\dfrac{(s-a)ds}{t},\ \hat{\omega}_2=\dfrac{(s-b)ds}{t}\ .
\end{eqnarray}
We assume \(a\neq b\) so that \(\hat{\omega}_1\) and \(\hat{\omega}_2\) becomes \(\mathbb{C}\)-linearly independent.\
If we introduce a constant matrix \(\Pi\) as 
\begin{eqnarray}
\Pi
:=2\pi i
\left( \begin{array}{cc} \int_{a_1}\hat{\omega}_1&\int_{a_2}\hat{\omega}_1\\ \int_{a_1}\hat{\omega}_2&\int_{a_2}\hat{\omega}_2\\ \end{array} \right)^{-1}, 
\end{eqnarray}
then, the normalized differentials of the first kind become
\begin{eqnarray}
\left( \begin{array}{c} \omega_1\\ \omega_2\\ \end{array} \right)
=\Pi\left( \begin{array}{c} \hat{\omega}_1\\ \hat{\omega}_2\\ \end{array} \right).
\end{eqnarray}
It should be noticed that the period matrix \(B\) becomes real in this setting.\ 
Moreover, thanks to the reciprocity law for compact Riemann surfaces, 
\begin{eqnarray}
&&\vec{U}_*=\left( \begin{array}{c} \int_{b_1}\chi_{P,Q}\\ \int_{b_2}\chi_{P,Q} \end{array} \right)
=\left( \begin{array}{c} \int_a^c\omega_1\\ \int_a^c\omega_2 \end{array} \right)
=\Pi\left( \begin{array}{c} \int_a^c\hat{\omega}_1\\ \int_a^c\hat{\omega}_2 \end{array} \right)
=\Pi\left( \begin{array}{c} 0\\ \int_a^c\hat{\omega}_2 \end{array} \right)\label{U*}
\end{eqnarray}
and
\begin{eqnarray}
&&\vec{U}_1=\left( \begin{array}{c} \int_{b_1}\chi_{P}^{(1)}\\ \int_{b_2}\chi_{P}^{(1)} \end{array} \right)
=\left( \begin{array}{c} \omega_1|_{a}\\ \omega_2|_{a} \end{array} \right)
=\Pi\left( \begin{array}{c} \hat{\omega}_1|_{a}\\ \hat{\omega}_2|_{a} \end{array} \right)
= \Pi\left( \begin{array}{c} 0\\ \hat{\omega}_2|_{a} \end{array} \right)\label{U1}
\end{eqnarray}
become purely imaginary vectors and 
\begin{eqnarray}
\vec{U}_*=\gamma\vec{U}_1\nonumber
\end{eqnarray}
holds for certain real constant \(\gamma\). 
We notice that \(\int_a^c\hat{\omega}_2\neq 0 \) since \(\hat{\omega}_1\) and \(\hat{\omega}_2\) are linearly independent.\ It also should be noticed that \(\hat{\omega}_2|_{a}\neq 0\) because \(a\neq b\).\ 
Hence \(\gamma\) is nonzero.\
Lastly, we have to show that \(\vec{U}_1,\vec{U}_2\) and \(\vec{U}_*\) are purely imaginary vectors.\ 
As for \(\vec{U}_1\) and \(\vec{U}_*\), these immediate follow from the expressions (\ref{U*}) and (\ref{U1}).\
Similarly, the reciprocity law \((\vec{U}_2)_j=(\partial_s\omega_j|_{a})/2\) also tells us that \(\vec{U}_2\) is a purely imaginary.\\ \hfill ({\it The proof of theorem 6 is finished.})\\[.2\baselineskip]
{\bf Proposition 8} (Reality of genus-two solutions)\\
The solution 
\begin{eqnarray}
u=i(U/2-2\kappa/\gamma)=
\dfrac{i}{2}\left(-\dfrac{F^-_x}{F^-}+\dfrac{F^+_x}{F^-}-\kappa\int_{Q_1}^{P_*}\chi_{1}-4\kappa/\gamma\right)
\nonumber
\end{eqnarray}
as in (\ref{sol-U}) or (\ref{sol-b}) is real valued if we chose \(X, \{a_1,a_2,b_1,b_2\}, P_0\) and \(P_*\) as those in the proof of theorem 6 and corollary 7.\\[.2\baselineskip] 
{\it Proof.} We already saw that \(F\) is a real valued functions for real \(x\) and \(t\).\ This implies that \(-F^-_x/F^-+F^+_x/F^+\) is purely imaginary.\ Hence, the task left is to show that \(\int_{Q_1}^{P_*}\chi_{1}\) is a real constant.\ 
There might be sophisticated ways to see this but we here use an explicit expression of \(\chi_{1}\) as follows.\
\begin{eqnarray}
\chi_{1}=
\partial_a\left(\dfrac{t+\sqrt{f(a)}}{2t(s-a)}\right)ds+
\sum_{j=1}^2\alpha_j\hat{\omega}_j.
\end{eqnarray}
\(\alpha_j,\ (j=1,2)\) are real constants chosen so as to eliminate the \(a-\)periods of the first term in the r.h.s.\ 
This expression tells us that \(\int_{Q_1}^{P_*}\chi_{1}\) is real valued if \(Q_1\) as well as \(P_*\) is on \({\cal W}\).\
And \(Q_1\) does lie on \({\cal W}\).\
Because an integral \(\int_{Q}^{P}\chi_{1}\ (Q, P\in{\cal W})\)  tends to \(\infty \) when \(Q\to W_s\), to \(-\infty \) when \(Q\to W_e\), and is analytic when \(Q\in {\cal W}\backslash\{W_s,W_e\}\), there must exist points that agree with the definition of \(Q_1\).\ \hfill 
\fbox{\rule{0em}{.6em}}\\[.2em]

\section{Numerical example}
\begin{figure}
\includegraphics[width=.8\linewidth]{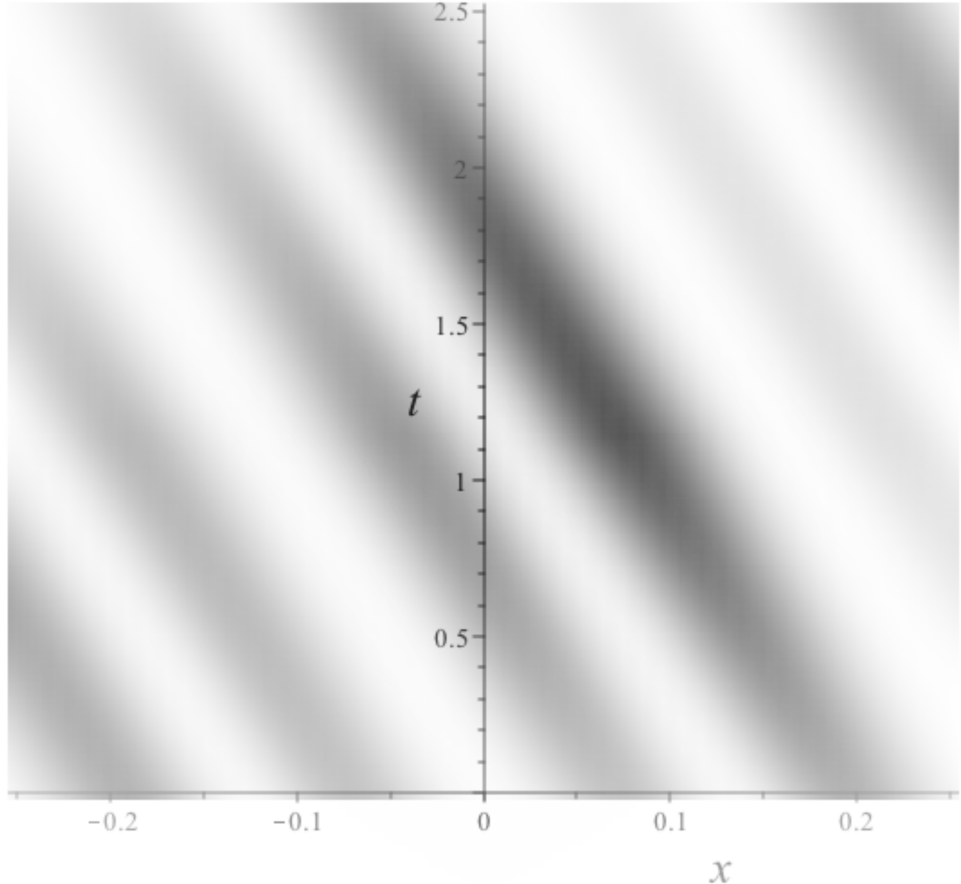}
\caption{\label{fig:TheSol} A density plot of the numerical solution \(u\) in section 7. \(-0.25<x<0.25,\ 0<t<2.\)}
\end{figure}

A numerical example of a genus-two theta function solution is given in this section.\
Here, Maple 2016 is used as the calculation software.\ 
All of the numbers given are rounded to four decimal places.\ 
We do not give these numbers with more accuracy than this, since the aim of this section is just to give a tangible image of what we have described in the previous sections, not to give precise shapes of 2-solitons.\
Suppose, for example, \(e_j=j\), then, the curve \(C\) namely becomes
\(t^2=(s-1)(s-2)(s-3)(s-4)(s-5)\).\ 
A numerical approximation of \(B\) matrix is;
\begin{eqnarray}
B= \left( \begin{array}{cc}  -7.8764&-3.1270\\ -3.1270&-6.2537\\ \end{array} \right).
\end{eqnarray}
We chose \(a=15\) and \(b=14.9995\) for the theorem 6.\
It should be noticed that we can make \(|\vec{U}_*|\) arbitrarily small by setting \(|a-b|\) small enough so that the analyticity condition of corollary 7 holds.\ Then, the transcendental equation becomes 
\begin{eqnarray}
\int_{15}^c\dfrac{(s-15)ds}{(s-1)(s-2)(s-3)(s-4)(s-5)}=0,
\end{eqnarray}
and a numerical  solution indicates \(c\approx 6.2152\).\
Then, we obtain
\begin{eqnarray}
\vec{U}_*=\left( \begin{array}{c} -0.6972i\\ 2.3808i\\ \end{array} \right),\ 
\vec{U}_1=\left( \begin{array}{c} - 9.6997i\\ 33.1222i\\ \end{array} \right),\
\vec{U}_2=\left( \begin{array}{c} - 1.0241i\\ -3.4981i\\ \end{array} \right),\
\gamma=0.0719.
\end{eqnarray}
If we set \(\kappa=i\) and \(\vec{Z}=0\), the relative depth \(\gamma\) becomes 55.6483 and \(U\) is now explicitly written as follows:
\begin{eqnarray}
&&U=\dfrac{f_x(x-0.0359i,t)}{f(x-0.0359i,t)}-\dfrac{f_x(x+0.0359i,t)}{f(x+0.0359i,t)},\\
&&f=\Theta\left(\left( \begin{array}{c} 9.6997\\ -33.1222\\ \end{array} \right)  x
+\left( \begin{array}{c}  1.0241\\ 3.4981\\ \end{array} \right)t\right).
\end{eqnarray} 
A density plot of the solution \(u\) for the ILW eqation (\ref{ILW-b}) is given in FIG.~\ref{fig:TheSol}.\ 

\section{Further discussion}
Genus-two solutions for the ILW equation are obtained in this paper.\ 
Following the method of this paper, one can obtain solutions of higher genus by finding colinear \(\vec{U}_*\) and \(\vec{U}_1\).\
However, it is probable that this would require another breakthrough in understanding.\ 
Another possible future work is to apply this technique to other nonlocal soliton equations such as the intermediate nonlocal nonlinear Schr\"{o}dinger equation.\ 
It should be noticed from the proof of the corollary 7 that the existence of the solution is only ensured for small \(|{\rm Im}(\gamma/\kappa)|\), which corresponds to the Benjamin-Ono limit.\ 
To construct a solution around the KdV limit is one more challenging problem.\
%
%
%
\begin{acknowledgments}
This work was supported by KAKENHI 26800064 
from the Japan Society for the Promotion of Science(JSPS).
\end{acknowledgments}

\begin{thebibliography}{9}
\bibitem{J77}
R. I. Joseph,
J.\ Phys.\ A:\ Math.\ Gen.
 {\bf10},
L225 (1977).
\bibitem{JE78}
R. I. Joseph and R. Egri,
J.\ Phys.\ A:\ Math.\ Gen. 
 {\bf11}, L97 (1978).
\bibitem{KK78}
T. Kubota and D. R. S. Ko, 
J.\ Hydronautics 
 {\bf12}, 157 (1978).
\bibitem{NM80}
A. Nakamura and Y. Matsuno, 
J.\ Phys.\ Soc.\ Japan 
{\bf 48}, 653 (1980).
\bibitem{P92}
A. Parker, 
J.\ Phys.\ A:\ Math.\ Gen. 
{\bf 25}, 2005 (1992).
\bibitem{CL79}
H. H. Chen and Y. C. Lee, 
Phys.\ Rev.\ Lett. 
{\bf 43}, 264 (1979).
\bibitem{M79}
Y. Matsuno, 
Phys.\ Lett. 
{\bf 74A}, 233 (1979).
\bibitem{KAS82}
Y. Kodama, M. J. Ablowitz and J. Satsuma, 
J.\ Math.\ Phys. 
{\bf 23}, 564 (1982).
\bibitem{Zai83}
A. Zaitsev, 
Sov.\ Phys.\ Dokl. 
{\bf 28}, 720 (1983).
\bibitem{Mil90}
T. Miloh, 
J.\ Fluid\ Mech. 
{\bf 211}, 617 (1990).
\bibitem{AFJS82}
M. J. Ablowitz, A. S. Fokas, J. Satsuma and H. Segar, 
J.\ Phys.\ A:\ Math.\ Gen. 
{\bf 15}, 781 (1982).
\bibitem{TS03}
Y. Tutiya and J. Satsuma, 
Phys.\ Lett.\ A 
{\bf 313}, 45 (2003). 
\bibitem{KSA81}
Y. Kodama and J. Satsuma and M. J. Ablowitz, 
Phys.\ Rev.\ Lett. 
{\bf 46}, 687 (1981). 
\bibitem{Ak61}
N. I. Akhiezer, 
Dokl.\ Acad.\ Nauk\ SSSR 
{\bf 2}, 687 (1961). 
\bibitem{Dub75}
B. A. Dubrovin, 
Funkt.\ Anal.\ Pril. 
{\bf 9(1)}, 63 (1975).
\bibitem{Dub75-2}
B. A. Dubrovin, 
Funkt.\ Anal.\ Pril. 
{\bf 9(3)}, 41 (1975).
\bibitem{IM75}
A. R. Its and V. B. Matveev, 
Teor.\ Mat.\ Fiz. 
{\bf 23(1)}, 51 (1975).
\bibitem{IM75-2}
A. R. Its and V. B. Matveev, 
Funkt.\ Anal.\ Pril. 
{\bf 9(1)}, 69 (1975).
\bibitem{MM75}
H. P. MacKean and P. van Moerbeke, 
Invent.\ Math. 
{\bf 30}, 217 (1975).
\bibitem{Kri76}
I. M. Krichever
Dokl.\ Acad.\ Nauk\ SSSR 
{\bf 227(2)}, 291 (1976).
\bibitem{Kri77}
I. M. Krichever, 
Funkt.\ Anal.\ Pril. 
{\bf 11(1)}, 15 (1977).
\bibitem{Kri81}
I. M. Krichever, 
{\it London Math. Soc. Lecture Notes Ser., vol. 60} (Cambridge University Press, Cambridge, 1981)
\bibitem{BBEIM94}
E. D. Belokolos and A. I. Bobenko and V. Z. Enol'skii and A. R. Its and V. B. Matveev, 
{\it Algebro-Geometric Approach to Nonlinear Integrable Equations} (Springer Verlag, Berlin Heiderberg, 1994)
\end{thebibliography}

\end{document}